\documentclass[a4paper,12pt]{article}
\pdfoutput=1 
\usepackage[english]{babel}
\usepackage{amsmath}
\usepackage{amssymb}
\usepackage{bbm}
\usepackage{color}
\usepackage{cite}
\usepackage{xcolor,hyperref}
\usepackage{graphicx}
\usepackage{xfrac}
\usepackage{enumitem}
\usepackage[titletoc,toc,title]{appendix}
\usepackage[font=footnotesize,labelfont=bf]{caption}
\Roman{section}
\newcommand{\naw}[1]{\left(#1\right)}
\newcommand{\ket}[1]{\left|#1\right>}

\newcommand{\av}[1]{\left<#1\right>}
\newcommand{\com}[1]{\left[#1\right]}
\newcommand{\modu}[1]{\left|#1\right|}

\title{Note on the Margolus-Levitin quantum speed limit for arbitrary fidelity}
\author{Krzysztof Andrzejewski$^1$\footnote{krzysztof.andrzejewski@uni.lodz.pl} \and Katarzyna Bolonek-Laso\'n$^2$\footnote{katarzyna.bolonek@uni.lodz.pl}  \and Piotr Kosi\'nski$^1$\footnote{piotr.kosinski@uni.lodz.pl} }
\date{%
	\small{	$^1$Department of Computer Science, Faculty of Physics and Applied Informatics University of Lodz, 149/153 Pomorska St., 90-236 Lodz, Poland\\$^2$Department of Statistical Methods, Faculty of Economics and Sociology\\ University of Lodz, 41/43 Rewolucji 1905 St., 90-214 Lodz,  Poland}}

\begin{document}
	\maketitle
	
	\begin{abstract}
		For vanishing fidelity between initial and final states two important quantum speed limits, the Mandelstam-Tamm limit (involving energy dispersion) and Margolus-Levitin one (involving excitation energy expectation value) have been derived. While the generalization of the former limit to the case of arbitrary fidelity is straightforward, the relevant generalization of the latter, given in the seminal paper by Giovanetti et al (\emph{Phys. Rev.} \textbf{A67} (2003), 052109) was based on the conjectured equality of lower and upper bounds on the right hand side of generalized Margolus-Levitin inequality, verified numerically up to seven digits. Only recently there appear two proofs of the conjecture. We provide below a very elementary new proof, based on the simplest tools from differential calculus. Thus the generalized Margolus-Levitin speed limit can be derived much in the spirit of the original one valid for vanishing fidelity.
		\end{abstract}
		
		\section{Inroduction}
		In recent decades much effort has been devoted to the study of the so called quantum speed limits, i.e.~a lower bounds on time it takes quantum system to evolve in some definite way.  They have been intensively studied in recent years since this problem is of great practical importance in quantum technologies. Many forms of quantum speed limits have been derived which involve various quantities: energy, fidelity, purity, entropy etc.~and concern closed (isolated) and open systems (see \cite{dodonov,frey,deffner} for reviews and \cite{mai,pati}, for more recent references). In what follows we will be considering the case of pure states of isolated systems.
		The prototype of quantum speed limit is the famous Mandelstam-Tamm \cite{Mandel} relation which gives the lower bound on time necessary for the quantum system to evolve from an initial state to the orthogonal one. It reads 
		\begin{equation}
			t\geq\frac{\pi}{2\Delta E},\label{a1}
		\end{equation}

	where 
	\begin{equation}
		\Delta E\equiv\sqrt{\av{\naw{\hat{H}-\big<\hat{H}\big>_0}^2}_0}
	\end{equation}	
is the energy dispersion in the initial state ($\big<\ldots\big>_0$ denotes here the expectation value in the initial state). Actually, a more general result can be derived \cite{Mandel,Bhatta,Pfeifer}: for
\begin{equation}
	\delta\equiv\modu{\av{\text{initial state}|\text{final state}}}^2
\end{equation}
	being the fidelity between the initial and final states, the evolution time is bounded by 	
		\begin{equation}
		t\geq\frac{\arccos\sqrt{\delta}}{\Delta E}.\label{a2}
	\end{equation}

\begin{figure}
	\centering
	\includegraphics[width=0.6\textwidth,keepaspectratio]{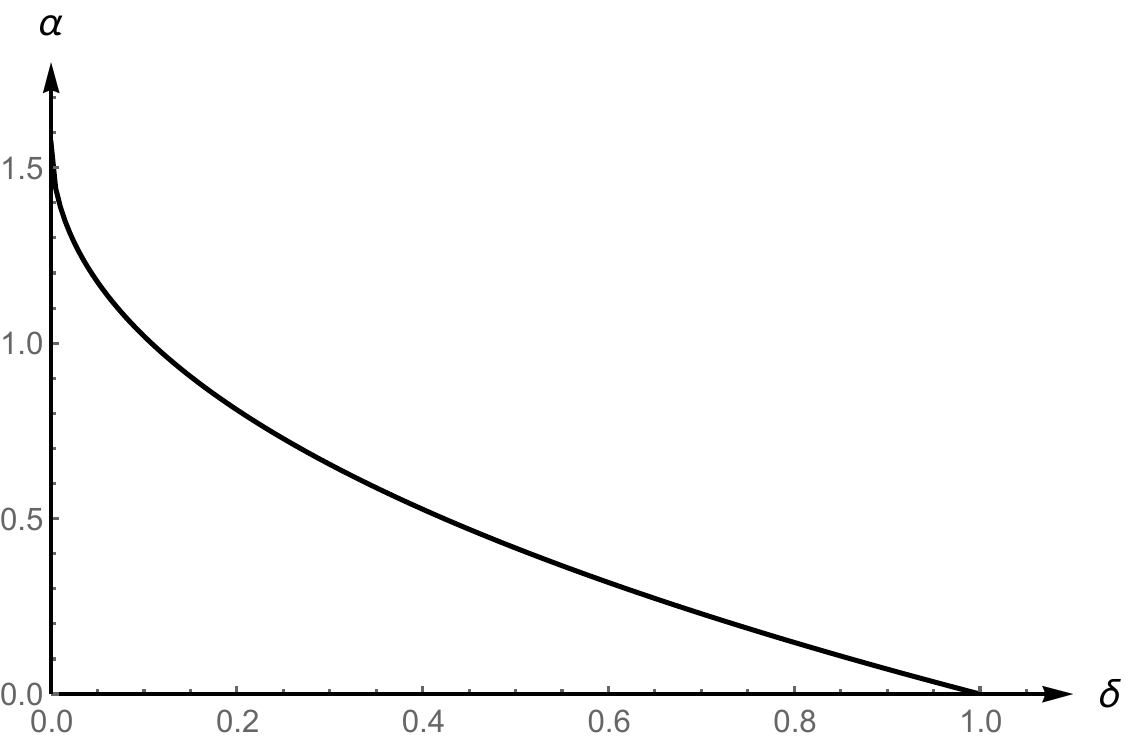}
	\caption{The function $\alpha(\delta)$ entering eq.~\eqref{a4}.}\label{f0}
\end{figure}

Margolus and Levitin \cite{Margolus} derived an alternative speed limit yielding lower bound for the orthogonalization time in terms of expectation value of energy in initial state. It reads
\begin{equation}
	t\geq\frac{\pi}{2\av{H-E_0}},\label{a3}
	\end{equation}	
with $E_0$ being the ground state energy of initial state. The natural question arises whether the inequality \eqref{a3} may be generalized to the case of arbitrary fidelity between the initial and the final state. Giovanetti et al.~\cite{Giovannetti} have shown that the relevant bound takes the form 
\begin{equation}
	t\geq\frac{\alpha(\delta)}{\av{H-E_0}},\label{a4}
\end{equation}
with $\alpha(\delta)$ being some function of fidelity. Although they did not provided the closed analytical formula for $\alpha$, the upper and lower bounds for the latter were given. Actually, these bounds agree numerically to 7 significant figures leading to the conjecture that, in fact, they coincide. If this is the case the exact formula for $\alpha$ is at our disposal.
 Its form is sketched on Fig.~\ref{f0}

Quite recently, H\"ornedal and S\"onnerborn \cite{Hornedal} provided a rigorous proof, based on symplectic geometry, that the exact formula for $\alpha$ coincides with the upper bound on it, derived in \cite{Giovannetti}. They rely heavily on symplectic geometric interpretation which suggests a specific character of Margolus-Levitin bound. The more advanced tools they use allow for drawing more detailed conclusions. Alternative proof of the coincidence of Giovannetti et al.~upper and lower bounds on $\alpha$ has been given in \cite{Chau}. It is, however, more involved and uses the arguments which seem to deviate significantly from the clear reasoning presented in Ref.~\cite{Giovannetti}.

In the present note we give a straightforward proof of the equality of upper and lower bounds on $\alpha$, derived in the paper by Giovannetti et al.~\cite{Giovannetti}.

\section{Lower and upper bounds coincide}

\begin{figure}[h!]
	\centering
	\includegraphics[width=0.6\textwidth,keepaspectratio]{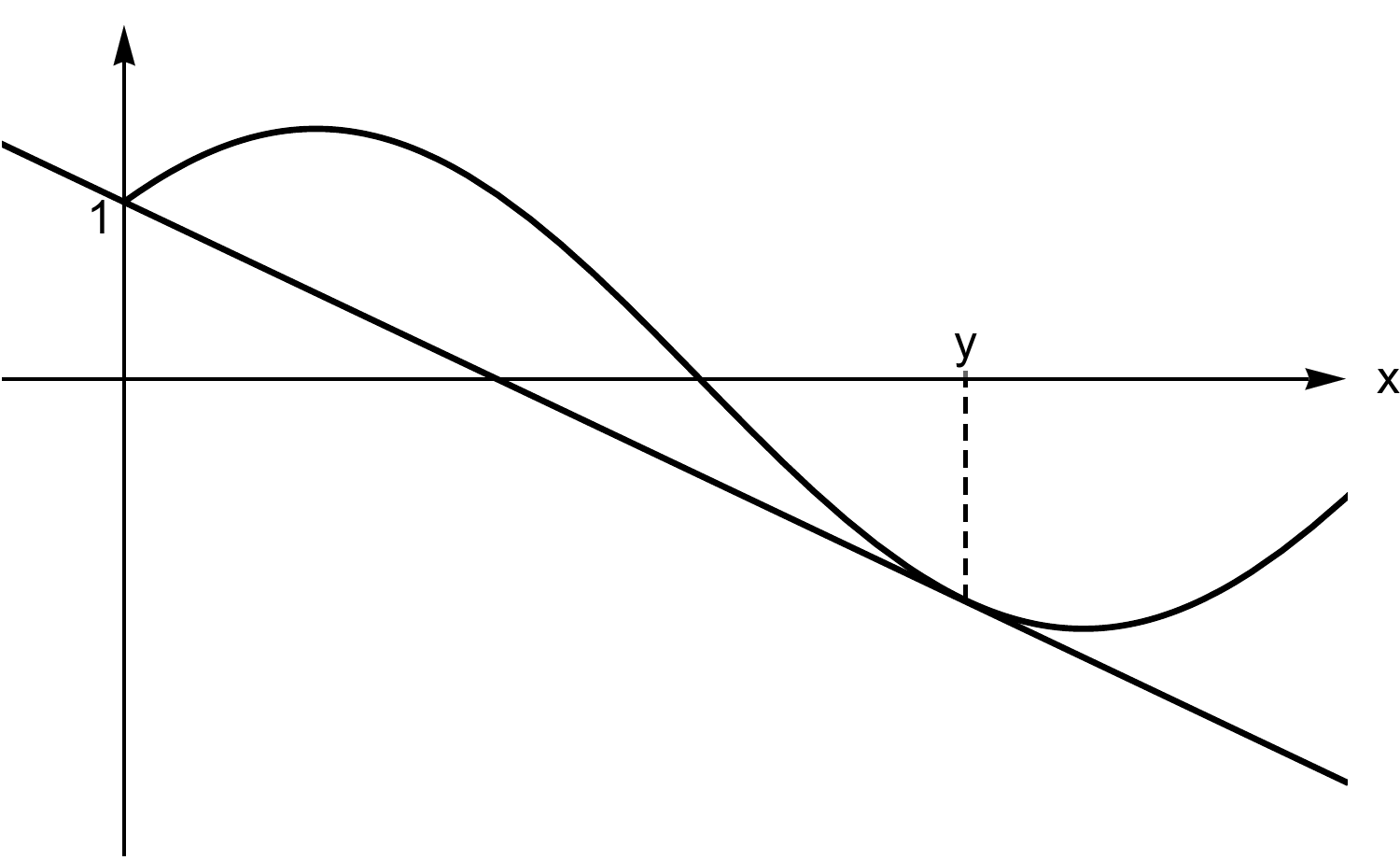}
	\caption{Configuration leading to optimal inequality \eqref{a6}.}\label{f1}
\end{figure}

	Giovanetti et al.~\cite{Giovannetti} derived the lower and upper bounds
	\begin{equation}
		m(\delta)\leq\alpha(\delta)\leq M(\delta)
	\end{equation}
on the function $\alpha(\delta)$ as follows. The upper bound results by considering the particular family of two-level states parametrized by one real variable $\xi\in[0,1]$,
\begin{equation}
	\ket{\Omega_\xi}=\sqrt{1-\xi^2}\ket{0}+\xi\ket{E_0}
\end{equation} 
where $\ket{0}$ and $\ket{E_0}$ are Hamiltonian eigenstates of energy 0 and $E_0$, respectively. By solving explicitly the dynamics one easily finds the evolution time as a function of fidelity between the initial and final state. Finally, by minimizing the latter function over $\xi\in[0,1]$ one finds the relevant upper bound.
 It reads
\begin{equation}
	M(\delta)=\frac{2}{\pi}\,\underset{z^2\leq\delta}{\min}\naw{\naw{\frac{1+z}{2}}\arccos\naw{\frac{2\delta-1-z^2}{1-z^2}}};\label{a5}
\end{equation} 
actually, the bound presented in \cite{Giovannetti} has a slightly different, but equivalent, form; we prefer the expression used in \cite{Hornedal}. The lower bound is obtained \cite{Giovannetti} by considering the family of inequalities 
\begin{equation}
	\cos x+q\sin x\geq 1-ax, \label{a6}
\end{equation}
for $x\geq 0$, $q\geq 0$. They are optimized (in the sense of choosing $a$ the smallest real number such that the inequality \eqref{a6} holds for all real $x\geq 0$)  by bounding the left hand side by the linear function tangent to it and equal to 1 for $x=0$ (cf.~Fig.~\ref{f1}).

Denoting by $y$ the $x$-coordinate of the tangent point one finds
\begin{align}
	\cos y+q\sin y&=1-ay,\label{a7}\\
-\sin y+q\cos y&=-a,\label{a8}
\end{align}
with $y$ being restricted to the interval 
\begin{equation}
	y\in \com{\pi-\arctan\bigg(\frac{1}{q}\bigg),\pi+\arctan(q)}.\label{a9}
\end{equation}
Eqs.~\eqref{a7}-\eqref{a9} define implicitly the function $a=a(q)$. Once this function is defined, the lower bound on $\alpha$ reads \cite{Giovannetti}
\begin{equation}
	m(\delta)=\frac{2}{\pi}\, \underset{0\leq\theta< 2\pi}{\min}\com{\underset{q\geq 0}{\max}\naw{\frac{1-\sqrt{\delta}(\cos\theta-q\sin\theta)}{a(q)}}}.\label{a10}
\end{equation}
Our aim here is to give a simple, straightforward proof of the equality 
\begin{equation}
	m(\delta)=M(\delta).\label{a11}
	\end{equation}

	We start with simplifying the notation in eq.~\eqref{a10} defining the lower bound. To this end let us put 
	\begin{align}
		&\rho\equiv\rho(\theta,\delta)\equiv 1-\sqrt{\delta}\cos\theta\geq 0,\label{b12}\\
		&\sigma\equiv\sigma(\theta,\delta)\equiv\sqrt{\delta}\sin\theta,\label{b13}\\
		& F(q,\theta,\delta)\equiv\frac{\rho+\sigma q}{a(q)}.\label{b14}
	\end{align}	
Then eq.~\eqref{a10} takes the form
\begin{equation}
	m(\delta)=\frac{2}{\pi}\, \underset{0\leq\theta<2\pi}{\min}\naw{\underset{q\geq 0}{\max}F(q,\theta,\delta)}.\label{b15}
\end{equation}
So, in order to prove the equality \eqref{a11} our only task is to analyze carefully the behaviour of $F(q,\theta,\delta)$ in the domain $q\geq 0$, $0\leq\theta<2\pi$, $0\leq\delta\leq 1$. Let us note that the main difficulty comes here from the denominator on the right hand side of eq.~\eqref{b14}; the function $a(q)$ is defined only implicitly by solving eqs.~\eqref{a7}$\div$\eqref{a9}. The main, yet very simple, idea to deal with this problem is to parametrize  the solutions to \eqref{a7} and \eqref{a8} in terms of the variable $y$. By solving them for $q$ and $a$ we find 
\begin{align}
	&q=\frac{1-\cos y-y\sin y}{\sin y-y\cos y},\label{b16}\\
	& a=\frac{1-\cos y}{\sin y-y\cos y},\quad y>0.\label{b17}
\end{align}
The function $a=a(q)$ is now given in parametrized form, $y$ being the relevant parameter.

\begin{figure}[h!]
	\centering
	\includegraphics[width=0.6\textwidth,keepaspectratio]{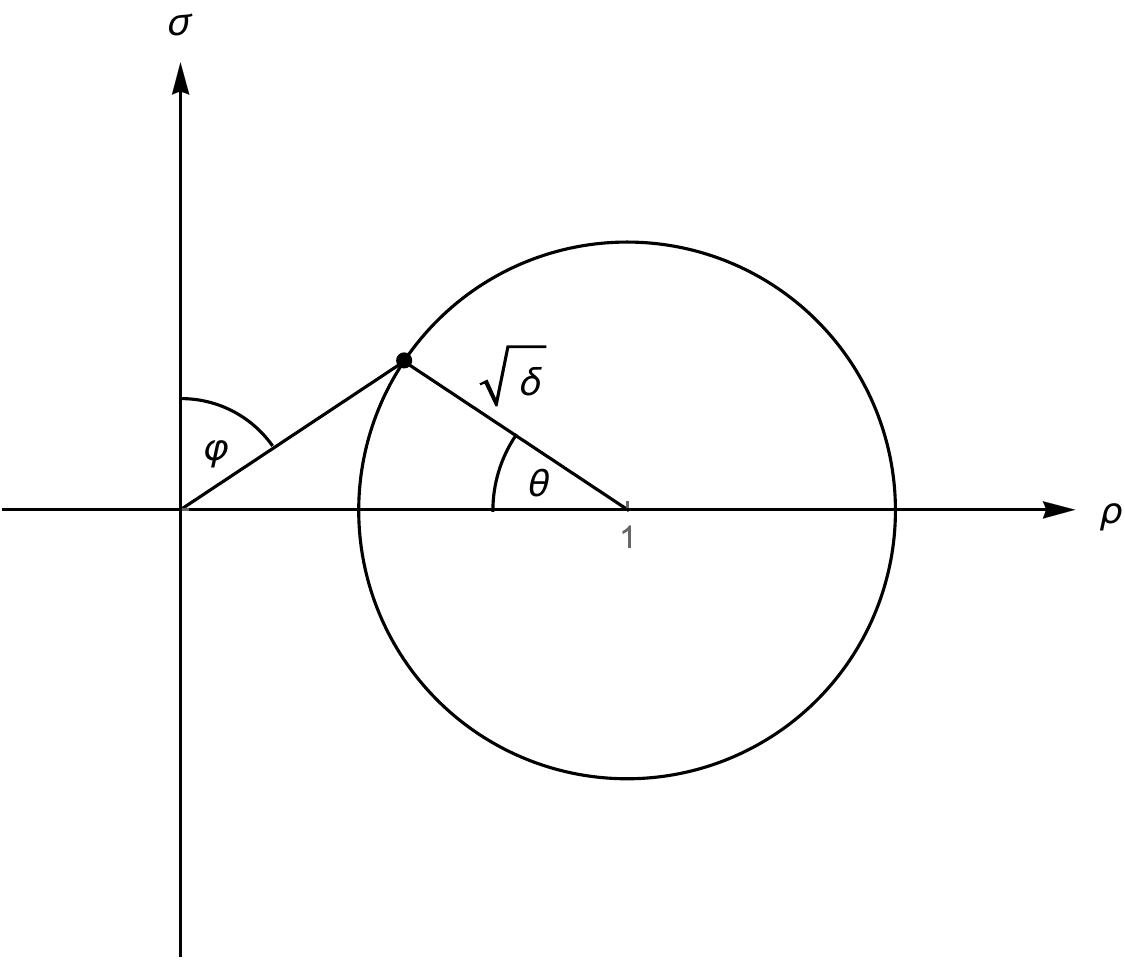}
	\caption{The set of points $(\rho,\sigma)$ defined by eqs.~\eqref{b12}, \eqref{b13}.}\label{f2}
\end{figure}

Before entering the detailed analysis of the function $a(q)$ let us make the following useful remarks. With $\delta$ fixed and $\theta$	varying in the interval $[0,2\pi)$ eqs.~\eqref{b12}, \eqref{b13} define a circle of radius $\sqrt{\delta}$, centered at $(0,1)$ in the $\rho-\sigma$ plane:
\begin{equation}
	(\rho-1)^2+\sigma^2=\delta.\label{b18}
\end{equation}
It appears to be convenient to define the angle $\varphi$ by
\begin{equation}
	\cos\varphi\equiv\frac{\sigma}{\sqrt{\rho^2+\sigma^2}},\quad\sin\varphi\equiv\frac{\rho}{\sqrt{\rho^2+\sigma^2}}\geq 0;\label{b19}
\end{equation}
then $\varphi$ belongs to some closed subinterval of $[0,\pi]$. The relevant geometry is depicted on Fig.~\ref{f2}

\begin{figure}[h!]
	\centering
	\includegraphics[width=0.6\textwidth,keepaspectratio]{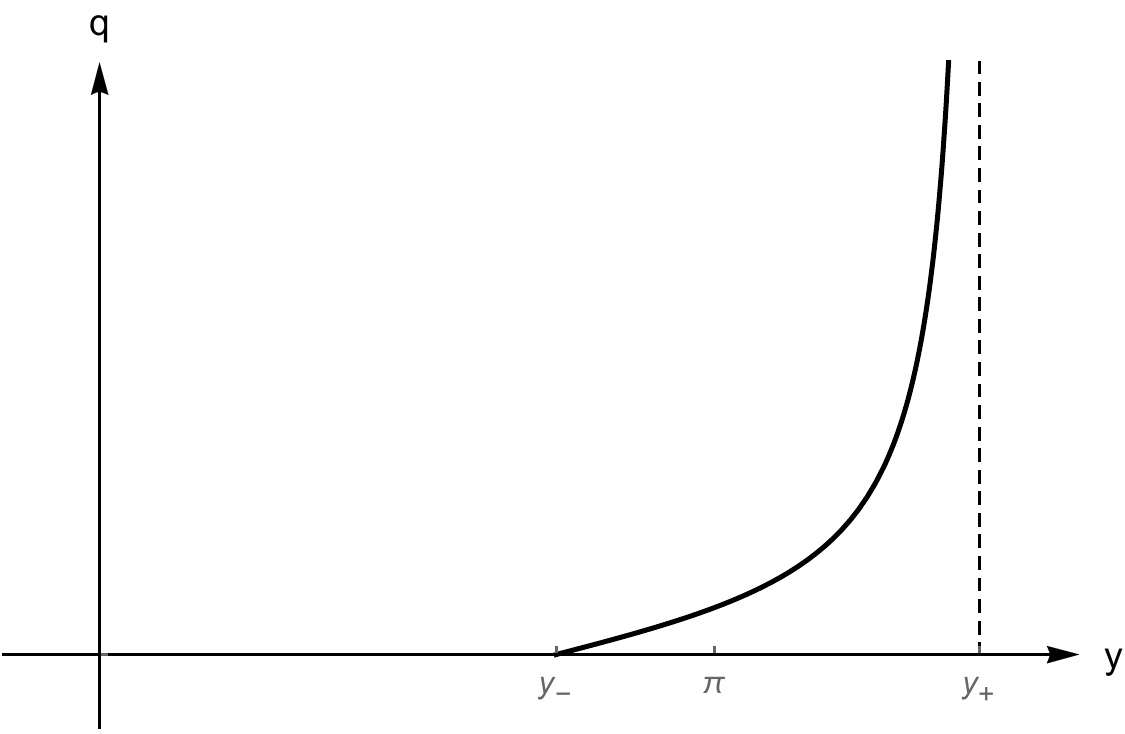}
	\caption{The function $q=q(y)$.}\label{f3}
\end{figure}

\begin{figure}
		\centering
	\includegraphics[width=0.6\textwidth,keepaspectratio]{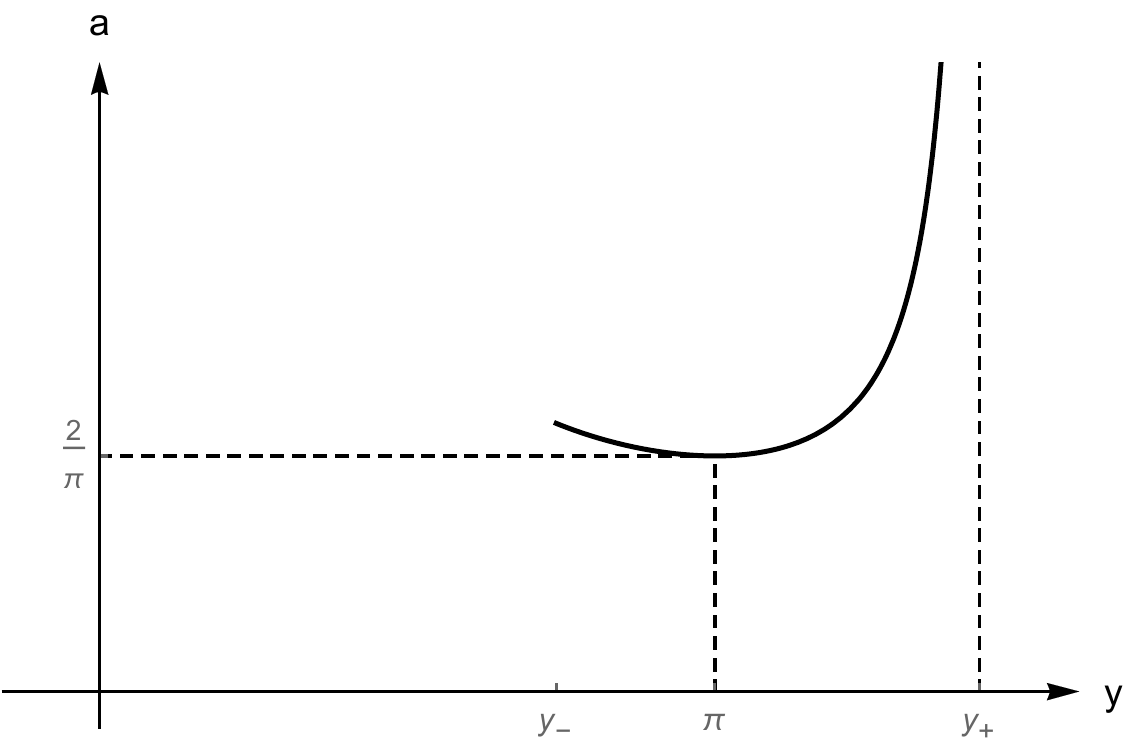}
	\caption{The function $a=a(y).$}\label{f4}
\end{figure}

Coming back to the analysis of the function $a=a(q)$, defined implicitly by eqs.~\eqref{b16}, \eqref{b17}, let us note that eq.~\eqref{b16} implies 
\begin{equation}
	\frac{dq}{dy}=\frac{y(y-\sin y)}{(\sin y-y\cos y)^2}>0,\label{b20}
\end{equation}

i.e.~$q$ is a monotonically growing function of $y$ wherever it is defined. We are interested in the solutions corresponding to $0\leq q<\infty$. It is easy to check that, due to the constraint \eqref{a9}, it is sufficient to consider the interval $y_-\leq y\leq y_+$, where $y_\pm$ are determined by the conditions
\begin{align}
	1-\cos y_--y_-\sin y_-&=0,\quad \frac{\pi}{2}<y_-<\pi,\label{b21}\\
	\sin y_+-y_+\cos y_+&=0, \quad \pi<y_+<\frac{3}{2}\pi.\label{b22}
\end{align} 
Numerically, $y_-=2.3311$, $y_+=4.4934$. In the interval $[ y_-,y_+)$ $q(y)$ is monotonically growing from 0 to $\infty$. Consequently, it is invertible. Inserting the function $y=y(q)$ into eq.~\eqref{b17} one obtains the relevant fuction $a=a(q)$. The functions $q=q(y)$, $a=a(y)$ on the interval $[y_-,y_+)$ are sketched on Fig.~\ref{f3} and \ref{f4}, respectively.

In order to analyze the function $a=a(q)$ note that 
\begin{equation}
	\frac{da}{dy}=\frac{\sin y(\sin y-y)}{(\sin y-y\cos y)^2}\label{b23}
\end{equation}
and
\begin{equation}
	\frac{da}{dq}=\frac{\sfrac{da}{dy}}{\sfrac{dq}{dy}}=-\frac{\sin y}{y}.\label{b24}
\end{equation}
In order to find the explicit form of $m(\delta)$ (cf.~eq.~\eqref{b15}) one has to find first the maximum of $F(q,\theta,\delta)$ for $q\in[0,\infty)$, $\theta$ and $\delta$ fixed; so we are looking for $\underset{q\geq 0}{\max}\, F(q,\theta,\delta)$.

According to the discussion above one can, equivalently, consider $F$ as the function of $y$ on the interval $[y_-,y_+)$. It reads 
\begin{equation}
	F(q(y),\theta,\delta)=\frac{\rho(\sin y-y\cos y)+\sigma(1-\cos y-y\sin y)}{1-\cos y} \label{b25}
\end{equation}
or, using \eqref{b12}, \eqref{b13} and \eqref{b19},
\begin{equation}
	F(q(y),\theta,\delta)=\sqrt{\rho^2+\sigma^2}\naw{\frac{\cos\varphi-\cos(\varphi+y)-y\sin(\varphi+y)}{1-\cos y}}.\label{b26}
\end{equation}
So we have to find the maximal value of $F(q(y),\theta,\delta)$ on the interval $[y_-,y_+)$. As always, the maximum may arise either at the internal point or the end point of the interval. In order to compute the relevant maximal value we find first the derivative 
\begin{equation}
	\frac{dF(q(y),\theta,\delta)}{dy}=\sqrt{\rho^2+\sigma^2}\cdot\frac{(y-\sin y)(\cos\varphi-\cos(\varphi+y))}{(1-\cos y)^2}.\label{b27}
\end{equation}
The maximal value can be attained in the internal point of $[y_-,y_+)$ only provided the derivative \eqref{b27} vanishes. This is possible iff
\begin{equation}
	\cos(\varphi+y)=\cos\varphi.\label{b28}
\end{equation}
Keeping in mind that  $0\leq\varphi\leq\pi$ and $\frac{\pi}{2}<y_-\leq y<y_+<\frac{3}{2}\pi<2\pi$ one finds the unique solution to eq.~\eqref{b28}
\begin{equation}
	y=2\pi-2\varphi.\label{b29}
\end{equation}
However, $y\in[y_-,y_+)$ yielding the following restrictions on $\varphi$
\begin{equation}
	\pi-\frac{1}{2}y_+<\varphi\leq\pi-\frac{1}{2}y_-.\label{b30}
\end{equation}

 The inequalities (\ref{b30}) provide the necessary conditions for $F(q(y), \theta,\delta)$ to attain the maximal value at the internal point of the interval $[y_-,y_+)$. On the contrary, if 
 $0\leq\varphi\leq\pi-\frac{1}{2}y_+$ or $\pi-\frac{1}{2}y_-<\varphi\leq\pi$, $\frac{dF(q(y),\theta,\delta)}{dy}$ does not vanish at the interval $[y_-,y_+)$ and attains its maximal value at one of its end points.
 
 The value of $\varphi$ depends, by virtue of eqs.~\eqref{b12}, \eqref{b13} and \eqref{b19}, on $\delta$ and $\theta$. Fixing $\delta$ and varying $0\leq\theta<2\pi$ one finds that $\varphi$ varies over some closed subinterval of $[0,\pi]$. There are a priori two possibilities:
 \begin{itemize}
 	\item[(i)] for all $\varphi$ the inequalities \eqref{b30} are obeyed. This is illustrated on Fig. \ref{f5}.

	\begin{figure}[h!]
	\centering
	\includegraphics[width=0.6\textwidth,keepaspectratio]{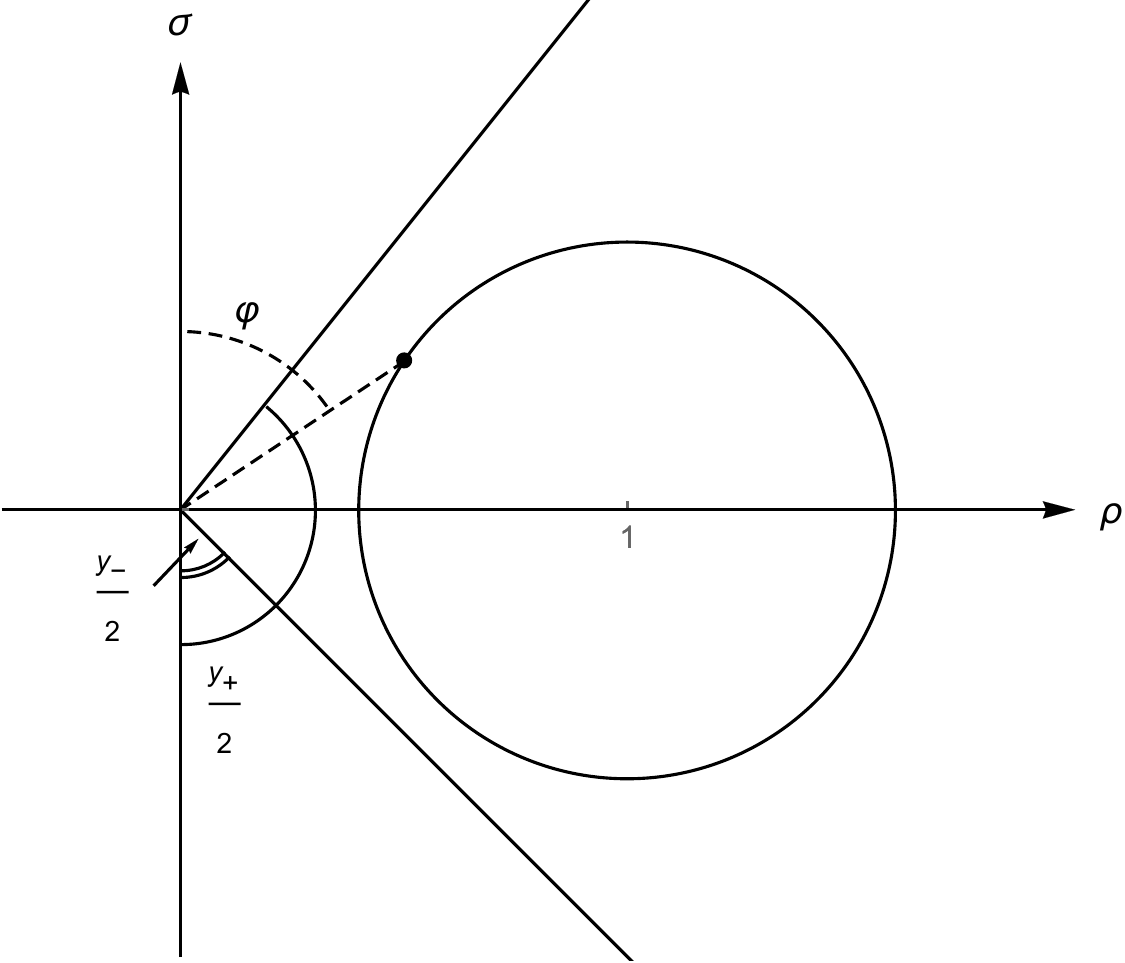}
	\caption{The inequalities \eqref{b30} are obeyed for all $(\rho,\sigma)$.}\label{f5}
\end{figure}
Then, for any $\varphi$ eq.~\eqref{b28} has the unique solution in the interval $[y_-,y_+)$. Moreover, one easily finds
\begin{equation}
	\frac{d^2F(q(y),\theta,\delta)}{dy^2}\bigg|_{F'(y)=0}=\frac{(y-\sin y)\sin(\varphi+y)}{(1-\cos y)^2}.\label{b31}
\end{equation}
Now, by virtue of \eqref{b29}, $\varphi+y=2\pi-\varphi$ so that $\pi+\frac{1}{2}y_+>\varphi+y\geq \pi+\frac{1}{2}y_-$ and $\sin(\varphi+y)<0$. Therefore, $F(q(y),\theta,\delta)$ attains unique maximum at $y=2\pi-2\varphi$. It equals
\begin{equation}
	F_{max}=\sqrt{\rho^2+\sigma^2}\naw{\frac{\pi-\varphi}{\sin\varphi}}\label{b32}
\end{equation}
or
\begin{equation}
	F_{max}=\naw{\frac{\rho^2+\sigma^2}{\rho}}\arccos\naw{\frac{-\sigma}{\sqrt{\rho^2+\sigma^2}}}.\label{b33}
\end{equation}
According to the equation \eqref{b15} the lower bound $m(\delta)$ is obtained by minimalizing $F_{max}$ over the points $(\rho,\sigma)$ running the circle $\eqref{b18}$. It is obvious that the minimum is attained for $\sigma\leq 0$. Defining
\begin{equation}
	\omega\equiv\sqrt{\delta}\cos\theta,\quad -\sqrt{\delta}\leq\omega\leq\sqrt{\delta},\label{b34}
\end{equation}
and taking into account the above remark allows us to write
\begin{equation}
	F_{max}=\naw{\frac{1-2\omega+\delta}{1-\omega}}\arccos\naw{\frac{\sqrt{\delta-\omega^2}}{\sqrt{1-2\omega+\delta}}}.
	\label{b35}
\end{equation} 
Using the identity $\arccos(2\tau^2-1)=2\arccos\tau$ we rewrite \eqref{b35} as
\begin{equation}
	F_{max}=\naw{\frac{1-2\omega+\delta}{2(1-\omega)}}\arccos\naw{\frac{\delta-1+2\omega-2\omega^2}{1-2\omega+\delta}}.
	\label{b36}
\end{equation}
Finally, we note that
\begin{equation}
	z=\frac{\delta-\omega}{1-\omega}\label{b37}
\end{equation}
defines one-to-one mapping of the interval $[-\sqrt{\delta},\sqrt{\delta}]$ onto itself. Eqs.~\eqref{b36}, \eqref{b37} imply
\begin{equation}
	F_{max}=\naw{\frac{1+z}{2}}\arccos\naw{\frac{2\delta-1-z^2}{1-z^2}}\label{b38} 
\end{equation}
yielding eq.~\eqref{a11}.
\item[(ii)] for some points on the circle $\eqref{b18}$ the inequalities \eqref{b30} are violated. This illustrated on Fig.~\ref{f6}. For any point $(\rho,\sigma)$ belonging either to the arc $AB$ or $CD$ the derivative $\frac{dF(q(y),\theta,\delta)}{dy}$ is nonvanishing in the interval $[y_-,y_+)$. Consider first the arc $AB$. Then $0\leq\varphi\leq\pi-\frac{1}{2}y_+$ and it is easy to see that $\frac{dF(q(y),\theta,\delta)}{dy}>0$ in the interval $[y_-,y_+)$. Therefore, the maximum is attained for $y\rightarrow y_+$ and reads
\begin{equation}
	F_{AB}=\frac{-\sigma}{\cos y_+}.\label{b39}
\end{equation}
 
On the contrary, on the arc $CD$ $\pi-\frac{1}{2}y_-<\varphi\leq\pi$ and $\frac{dF(q(y),\theta,\delta)}{dy}<0$ on the whole interval $[y_-,y_+)$. Therefore, the maximum is attained for $y=y_-$ and reads:
\begin{equation}
	F_{CD}=\frac{\alpha}{\sin y_-}.\label{b40} 
\end{equation}
\begin{figure}
	\centering
	\includegraphics[width=0.6\textwidth,keepaspectratio]{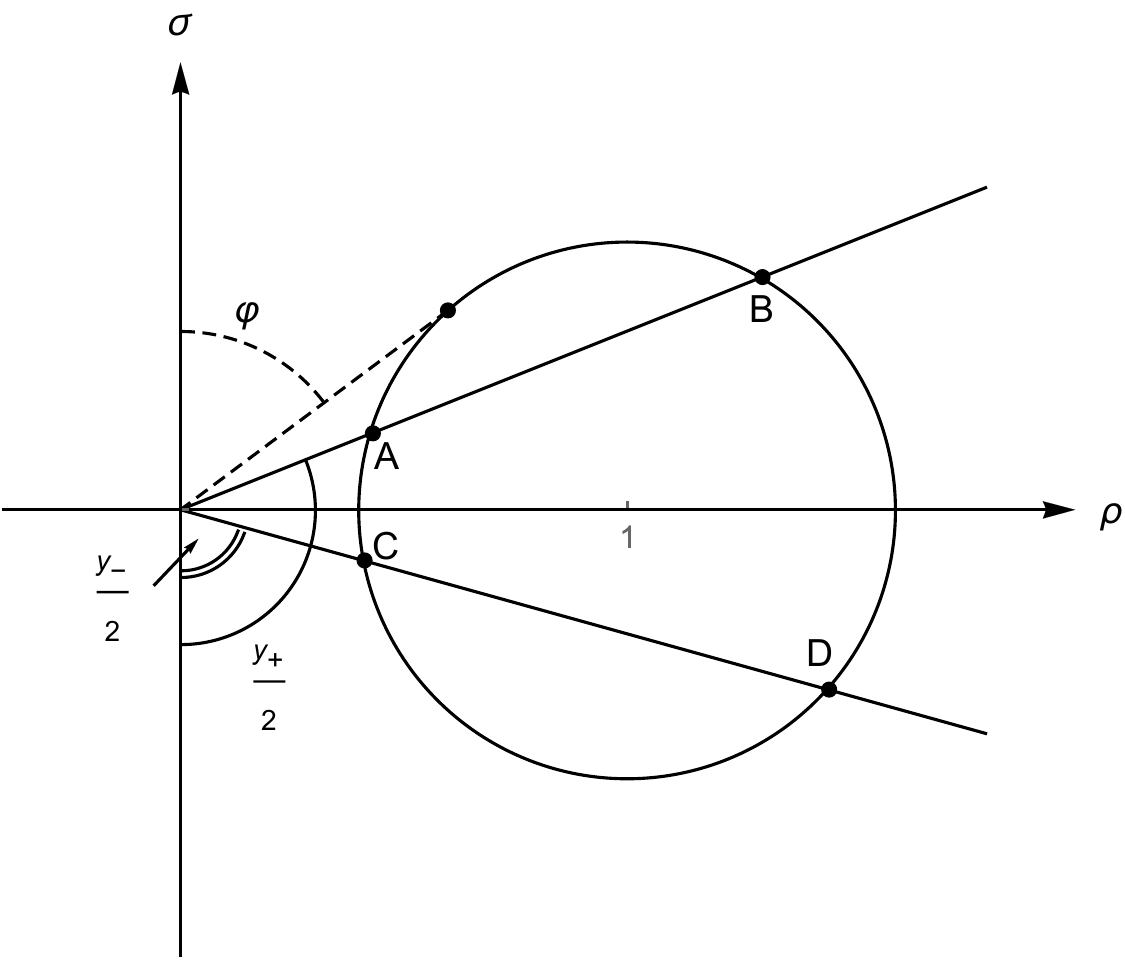}
	\caption{Geometric setting for violating the inequalities \eqref{b30}.}\label{f6}
\end{figure}
On the remaining arcs $AC$ and $BD$ the inequalities \eqref{b30} do hold and the reasoning leading to the eq.~\eqref{b33} remains valid. In order to find the minimal value of $\underset{q\geq 0}{\max}\,F(q,\theta,\delta)$ over the circle \eqref{b18} we have to compare $F_{AB}$, eq.~\eqref{b39}, $F_{CD}$, eq.~\eqref{b40} and $F_{max}$, eq.~\eqref{b33}. To this end we define $\psi\equiv\pi-\varphi$ and parametrize the points $(\rho,\sigma)$ by $\psi$ (in general, to any $\psi$ there correspond two points on the circle). The relevant parametrization of the circle \eqref{b18} is presented on Fig.~\ref{f7}.
\begin{figure}
	\centering
	\includegraphics[width=0.6\textwidth,keepaspectratio]{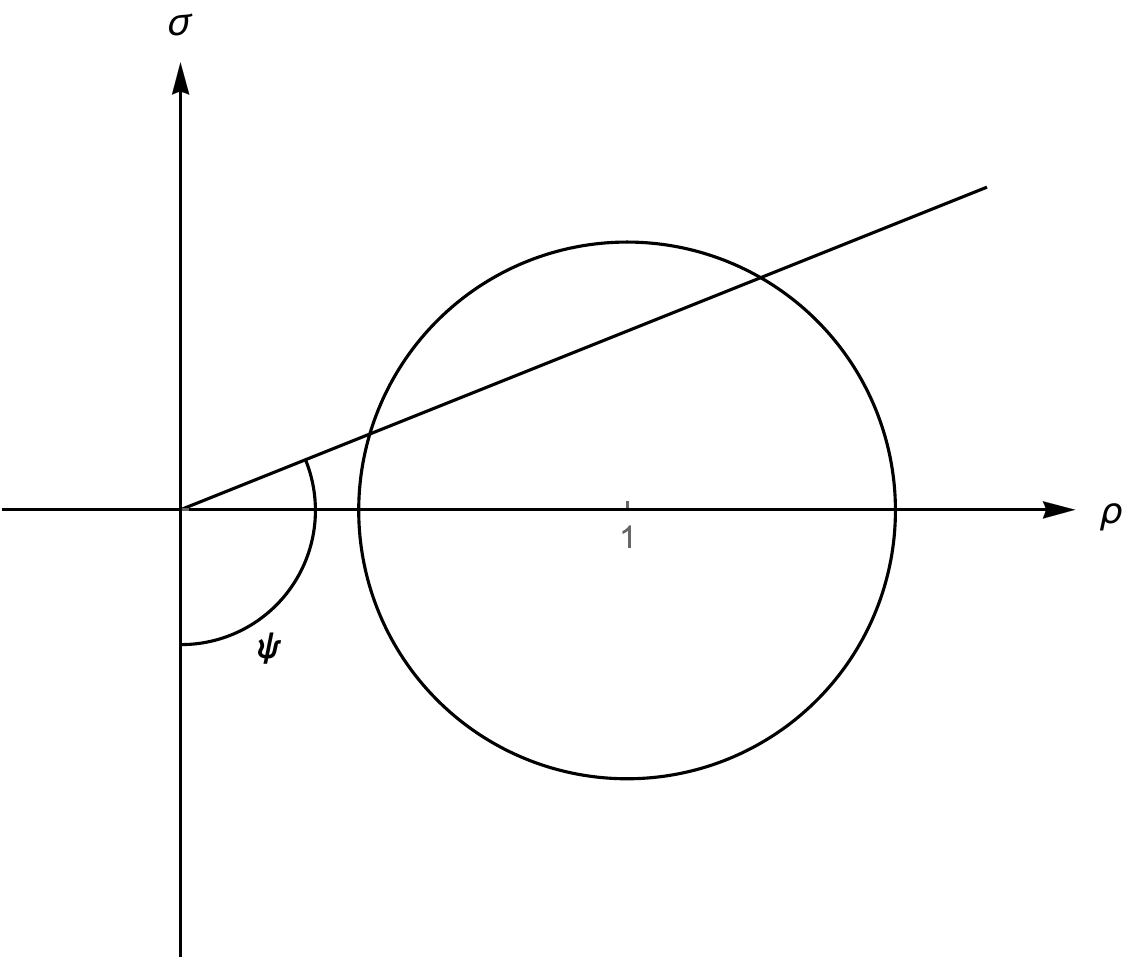}
	\caption{Parametrization of the circle \eqref{b18}.}\label{f7}
\end{figure}

The coordinates $(\rho,\sigma)$ of two intersection points obey the equations
\begin{align}
	& \rho+\sigma\tan\psi=0,\label{a41}\\
	& (\rho-1)^2+\sigma^2-\delta=0,\label{a42}
\end{align}
which yield
\begin{align}
	&\rho=\sin^2\psi\pm\sin\psi\sqrt{\delta-\cos^2\psi},\label{a43}\\
	&\sigma=-\sin\psi\cos\psi\mp\cos\psi\sqrt{\delta-\cos^2\psi}.\label{a44}
	\end{align}
Then
\begin{align}
	&F_{max}=\naw{2+\frac{\delta-1}{\sin^2\psi\pm\sin\psi\sqrt{\delta-\cos^2\psi}}}\psi,\label{a45}\\
	&F_{AB}=\frac{\sin\psi\cos\psi\pm\cos\psi\sqrt{\delta-\cos^2\psi}}{\cos y_+},\label{a46}\\
	&F_{CD}=\frac{\sin^2\psi\pm\sin\psi\sqrt{\delta-\cos^2\psi}}{\sin y_-}.\label{a47}
\end{align}
Now, it is straightforward to check that 
\begin{equation}
	F_{max}-F_{AB}=\frac{\sin\psi\pm\sqrt{\delta-\cos^2\psi}}{2\sin\psi}\naw{2\psi-\frac{\sin 2\psi}{\cos y_+}}.\label{a48}
\end{equation}
On the $AB$ arc $2\pi\geq2\psi\geq y_+>\pi$ and the right hand side is nonnegative; it vanishes for $2\psi=y_+$ which corresponds to the points $A$ and $B$. Moreover $F_{AB}$ takes the minimal value at $A$. As a result, $F_{max}$ and $F_{AB}$ attain the same minimal value on the $AB$ arc.

Analogously,
\begin{equation}
	F_{max}-F_{CD}=\frac{\sin\psi\pm\sqrt{\delta-\cos^2\psi}}{2\sin\psi}\naw{2\psi-\frac{1-\cos 2\psi}{\sin y_-}}.\label{a49}
\end{equation}
On the $CD$ arc $0\leq 2\psi\leq y_-$ and the right hand side is nonnegative; it vanishes for $2\psi=y_-$ which corresponds to the points $C$ and $D$. $F_{CD}$ takes the minimal value at $C$. Again we conclude that $F_{max}$ and $F_{CD}$ attain the same minimal value on the $CD$ arc.
\end{itemize} 
It follows from the above discussion that also in the (ii) case one obtains the correct value of $m(\delta)$ by minimalizing $F_{max}$ on the whole circle \eqref{b14}. Therefore, we can refer to the reasoning described in (i). This concludes the proof.


\begin{thebibliography}{99}
	
		\bibitem{dodonov} V. Dodonov, A. Dodonov, Energy-time and frequency-time uncertainty relations: exact inequalities, \emph{Phys. Scr.} \textbf{90} (2015), 074049
		\bibitem {frey} M. Frey, Quantum speed limits-primer, perspectives, and potential future directions, \emph{Quant. Inf. Proc.} \textbf{15} (2016), 3919
		\bibitem{deffner} S. Deffner, S. Campbell, Quantum speed limits: from Heisenberg's uncertainty principle to optimal quantum control, \emph{J. Phys. A: Math. Theor.} \textbf{50} (2017), 453001
		\bibitem{mai} Z. Mai, C. Yu, Tight and attainable quantum speed limit for open systems, \emph{Phys. Rev.} \textbf{A108} (2023), 052207
		\bibitem{pati} A. Pati, B. Mohan, S. Braunstein, Exact quantum speed limits, arXiv: 2305.038339
		\bibitem{Mandel} L. Mandelstam, I. Tamm, The uncertainty relation between energy and time in nonrelativistic quantum mechanics,  \emph{Journ. Phys} \textbf{9} (1945), 249
		\bibitem{Bhatta} K. Bhattacharyya, Quantum decay and the Mandelstam-Tamm-energy inequality, \emph{Journ, Phys.} \textbf{A16} (1983), 2993
		\bibitem{Pfeifer} P. Pfeifer, How fast can a quantum state change with time?, \emph{Phys. Rev. Lett.} \textbf{70} (1993), 3365
		\bibitem{Margolus} N. Margolus, L.B. Levitin, The maximum speed of dynamical evolution, \emph{Physica} \textbf{D120} (1998), 188
		\bibitem{Giovannetti} V. Giovannetti, S. Lloyd, L. Maccone, Quantum limits to dynamical evolution, \emph{Phys. Rev.} \textbf{A67} (2003), 052109
		\bibitem{Hornedal} N. H\"ornedal, O. S\"onnerborn, The Margolus-Levitin quantum speed limit for an arbitrary fidelity, arXiv: 2301.10063
		\bibitem{Chau} H.F. Chau, $\alpha_>(\varepsilon)=\alpha_<(\varepsilon)$ for the Margolus-Levitin quantum speed limit bound, arXiv: 2305.10101
	\end{thebibliography}
\end{document}